\documentclass[sigconf]{acmart}

\AtBeginDocument{%
  \providecommand\BibTeX{{%
    \normalfont B\kern-0.5em{\scshape i\kern-0.25em b}\kern-0.8em\TeX}}}

\usepackage{comment}
\usepackage{balance}
\usepackage{hyperref}

\setcopyright{iw3c2w3}

\begin{document}

\copyrightyear{2020}
\acmYear{2020} 
\acmConference[WWW '20 Companion]{Companion Proceedings of the Web Conference 2020}{April 20--24, 2020}{Taipei, Taiwan}
\acmBooktitle{Companion Proceedings of the Web Conference 2020 (WWW '20 Companion), April 20--24, 2020, Taipei, Taiwan}
\acmPrice{}
\acmDOI{10.1145/3366424.3383567}
\acmISBN{978-1-4503-7024-0/20/04}

\newcommand{\benj}{\color{red}}
\newcommand{\mizvol}{\color{blue}}

\title{%
    What is Trending on Wikipedia? \\
    Capturing Trends and Language Biases Across Wikipedia Editions}
    
\author{Volodymyr Miz, Jo\"{e}lle Hanna, Nicolas Aspert, Benjamin Ricaud, and Pierre Vandergheynst}
\affiliation{\institution{LTS2, EPFL, Station 11, CH-1015 Lausanne, Switzerland}}
\email{firstname.lastname@epfl.ch}

\renewcommand{\shortauthors}{Volodymyr Miz, et al.}

\begin{abstract}


In this work, we propose an automatic evaluation and comparison of the browsing behavior of Wikipedia readers that can be applied to any language editions of Wikipedia. As an example, we focus on English, French, and Russian languages during the last four months of 2018. The proposed method has three steps. Firstly, it extracts the most trending articles over a chosen period of time. Secondly, it performs a semi-supervised topic extraction and thirdly, it compares topics across languages.
The automated processing works with the data that combines Wikipedia's graph of hyperlinks, pageview statistics and summaries of the pages.

The results show that people share a common interest and curiosity for entertainment, e.g. movies, music, sports independently of their language. Differences appear in topics related to local events or about cultural particularities. Interactive visualizations showing clusters of trending pages in each language edition are available online \hyperlink{https://wiki-insights.epfl.ch/wikitrends}{https://wiki-insights.epfl.ch/wikitrends}
\end{abstract}



\maketitle

\section{Introduction}
Wikipedia has become one of the most popular sources of information in the world. As of January 2020, it is written in 309 languages. Wikipedia readers have different motivations~\cite{singer2017we}. The range and depth of their interests are very diverse. Some readers may look for an up-to-date source of information related to an event that appeared in the news, while others are interested in solving work or school-related tasks. Some may be satisfied with a quick overview of a topic, while others strive for an in-depth understanding of all related facts.


\begin{figure*}
  \includegraphics[width=\textwidth, trim={0cm 5.5cm 0cm 5.5cm}, clip]{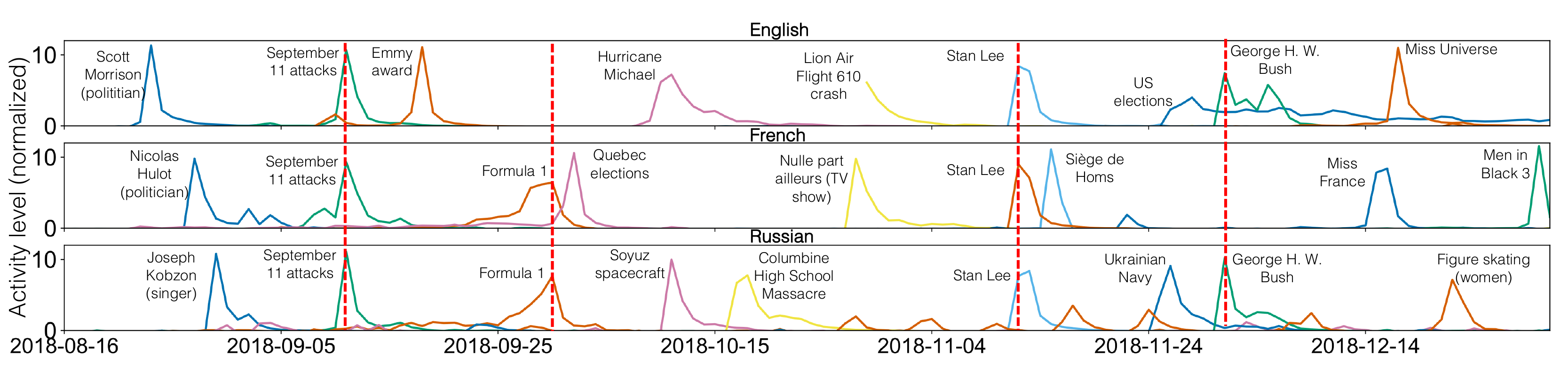}
  \caption{Trends over time and across languages. Each colored curve, together with a short description, is associated to a trend. They represent the number of visits on the pages belonging to the trend over time. The associated keyword is the title of the most central page of the trend. Red dashed lines highlight the moments when readers' interests in multiple language editions coincide.}\label{fig:teaser_timeseries}
\end{figure*}

Due to these features, Wikipedia has become a global and unified information medium for people having different backgrounds, hobbies, religious affiliations, political views, and languages. This makes Wikipedia an open window on cultural differences across different languages and populations. Analysing Wikipedia’s viewership statistics could help identify collective interests of socially diverse communities of people speaking different languages and compare them over chosen periods of time.

The structure and the content of the online encyclopedia differ across languages. Some editions are more developed than others. When it comes to the coverage of certain topics, the level of detail and the content varies largely from one language to another. This difference is especially noticeable when we look at controversial articles related to culture, politics or history, as it is shown in~\cite{russell2019joy} for the web in general and in~\cite{hecht2010tower,bao2012omnipedia} for Wikipedia in particular. For instance, the Italian version of the article about Leonardo da Vinci is much more detailed than the one in English. The content difference is also apparent in fairly non-controversial topics. For example, the article "Cat" in Spanish focuses on the animal's diseases and covers this aspect in greater detail than the same article written in English. In addition to the different texts, different hyperlinks are also present pointing to different concepts or related pages ~\cite{hecht2010tower}. Such differences may trigger diverging associations depending on the language in which one reads the very same article. 

These observations lead to an interesting question: is it possible to quantify differences between languages automatically?

In this work, we identify and compare the most popular topics among readers of three Wikipedia language editions, English, French, and Russian, over the last 4 months of 2018. We focus on articles that have unusual spikes in the number of views and associate those spikes of activity with real-world events. We discuss similarities and differences in the preferences of the readers of the three editions.

We define trends as popular topics having a rise of interest from readers during a limited period of time. We rely on the bursty behavior of page views, noticed and modeled in previous works~\cite{ratkiewicz2010characterizing,thij2012modeling,miz2019anomaly}. We require the trend to correspond to a group of Wikipedia articles connected by hyperlinks and having a correlated increase in viewership. This allows discarding isolated pages that get a high number of hits due to 'flash crowd' coming from a popular page pointing to a Wikipedia article, which is a very common use case. We rely on the method presented in~\cite{miz2019anomaly} to extract the trends. From this method, we obtain well-separated clusters of Wikipedia pages and their number of visits over time. This trend extraction is illustrated on Fig.~\ref{fig:teaser_timeseries} where the most trending events and their popularity are displayed on a timeline.

Our first contribution is a flexible method that automatically: 
\begin{itemize}
    \item extracts the most popular topics during a chosen period of time (trends)
    \item labels each trend according to the summaries of Wikipedia pages related to it.
\end{itemize}
The method is language-independent and relies solely on Wikipedia pageview data and graph of hyperlinks. 

Our second contribution is the analysis of trends, their similarities and differences across three language editions of Wikipedia.

\section{Related work}

The first studies on the interest of readers~\cite{lehmann2014reader,spoerri2007popular} used Wikipedia page view counts to identify the most popular articles. They found out that entertainment (movies, music, sports, etc.) and people's biographies are the most popular topics that interest Wikipedia readers. Recently, a more advanced method based on the dynamic evolution of popularity~\cite{miz2019anomaly} confirmed this tendency and shed light on the relationship between events that appear in the news, trends and Wikipedia article popularity.

Another topic of recent investigations is the motivation of Wikipedia readers~\cite{singer2017we}. It was reported that among the main motivations for reading about a particular topic is that it was referenced in the media (30\%), in a conversation (22\%) or it is a current event (13\%). This indicates a strong influence of trends and news on the readers' consumption of articles. Trends are an essential part of the search for information and that is what we want to extract and analyse. They are highly influenced by the readers' environment and hence should reflect similarities and differences across languages and cultures. However, the previously cited works are restricted to the English version of Wikipedia.

Concerning language biases and differences between versions of Wikipedia, there are two distinct groups of studies related to them. The first group studies editorial activity and page content while the second focuses on the viewership and readers' behavior.

Editorial differences across languages have been investigated for particular topics such as medicine-related articles~\cite{domingues2019characterizing}, aircraft crashes~\cite{garcia2016dynamics} or edit wars on popular pages~\cite{yasseri2014most}. They all point out differences due to political and cultural influences associated to languages. Depending on the language some parts within an article are more detailed, biased or more debated.
As a consequence of this editorial behavior, page contents in different languages contain noticeable variations when comparing articles about famous people\cite{callahan2011cultural}, food-related pages~\cite{laufer2015mining} or history of states~\cite{samoilenko2017analysing}.
Not only the text is different but also the hyperlink structure~\cite{gabella2018cultural}.

With the help of a large-scale poll, Lemmerich et al.~\cite{lemmerich2019world} analyzed visitors' motivations across 14 languages. The authors demonstrated that Wikipedia viewership patterns and use cases vary in different language editions and connected their findings to the socio-economic characteristics in certain countries, such as the Human Development Index.

To our knowledge, this is the first study focusing on the automatic detection of viewership trends and biases in multiple language editions of Wikipedia. In contrast to previous works, topics are not predefined but extracted automatically according to their popularity (number of visits on the related pages). We provide an overview of the dynamical evolution of the most popular topics among the readers of English, French, and Russian language editions of Wikipedia and reveal differences and commonalities between them.


\section{Dataset}
In this work, we use the graph-structured dataset for Wikipedia research described in~\cite{aspert2019graph} where nodes represent Wikipedia articles and edges correspond to hyperlinks between them. The associated open source repository provides tools for pre-processing the official Wikipedia SQL dumps (hyperlinks and pageviews records) in all available languages. 

We studied English, French, and Russian language editions of Wikipedia. Each language edition was considered separately and we did not take into account inter-language hyperlinks.

The dataset also contains hourly viewership activity data. We pre-processed each Wikipedia dumps throughout August-December 2018. The total number of remaining articles after the pre-processing is about 60 000. Details of the pre-processing are described in Section~\ref{sec:method}.

We decided to focus specifically on three languages of Wikipedia for two reasons. First and the foremost, the availability of native speakers who served as experts while assessing the quality of topic detection. Second, the size and popularity of Wikipedia editions. The three chosen Wikipedia editions are relatively large, popular, and have a well-developed hyperlink infrastructure.

We did not use any location information and considered only the language of Wikipedia articles. Therefore, we do not attribute our results to specific locations. The results of the study reflect only biases among the readers of different language editions of Wikipedia who may be located in different parts of the world.

\section{Method}\label{sec:method}

Our approach consists of the following steps:
\begin{enumerate}
    \item Extraction of a sub-network of trending Wikipedia articles and identification of trends.
    \item Extraction of keywords from the summaries of every Wikipedia article in the sub-network and weighting according to their importance.
    \item Labeling of the trends with high-level topics using the extracted keywords.
\end{enumerate}

\textbf{Step (1): Wikipedia sub-network of trends.} 
To extract sub-networks that comprise trending Wikipedia articles, we use a graph-based anomaly detection algorithm presented in~\cite{miz2019anomaly}. The algorithm extracts a sub-network from the Wikipedia network of pages and hyperlinks. From the initial graph, it keeps the pages that encounter surges of user interest (viewership bursts) over time, as well as their connections. A viewership burst corresponds to a spike in the activity of a page. The desired magnitude of a spike is controlled by a hyperparameter in the algorithm. Besides, the algorithm sets a weight on each hyperlink connection. This weight is increased for pages having a correlated bursty behavior, reflecting their correlation strength. This automatically creates clusters of trending, correlated, pages.
These clusters of densely connected pages within the sub-network are extracted using a community detection algorithm~\cite{blondel2008fast}. We may assume that every cluster of pages is related to an event or a topic that attracted the interest of Wikipedia visitors at a certain moment in time. 

\begin{figure*}
  \includegraphics[width=\textwidth, trim={0cm 0cm 0cm 0cm}, clip]{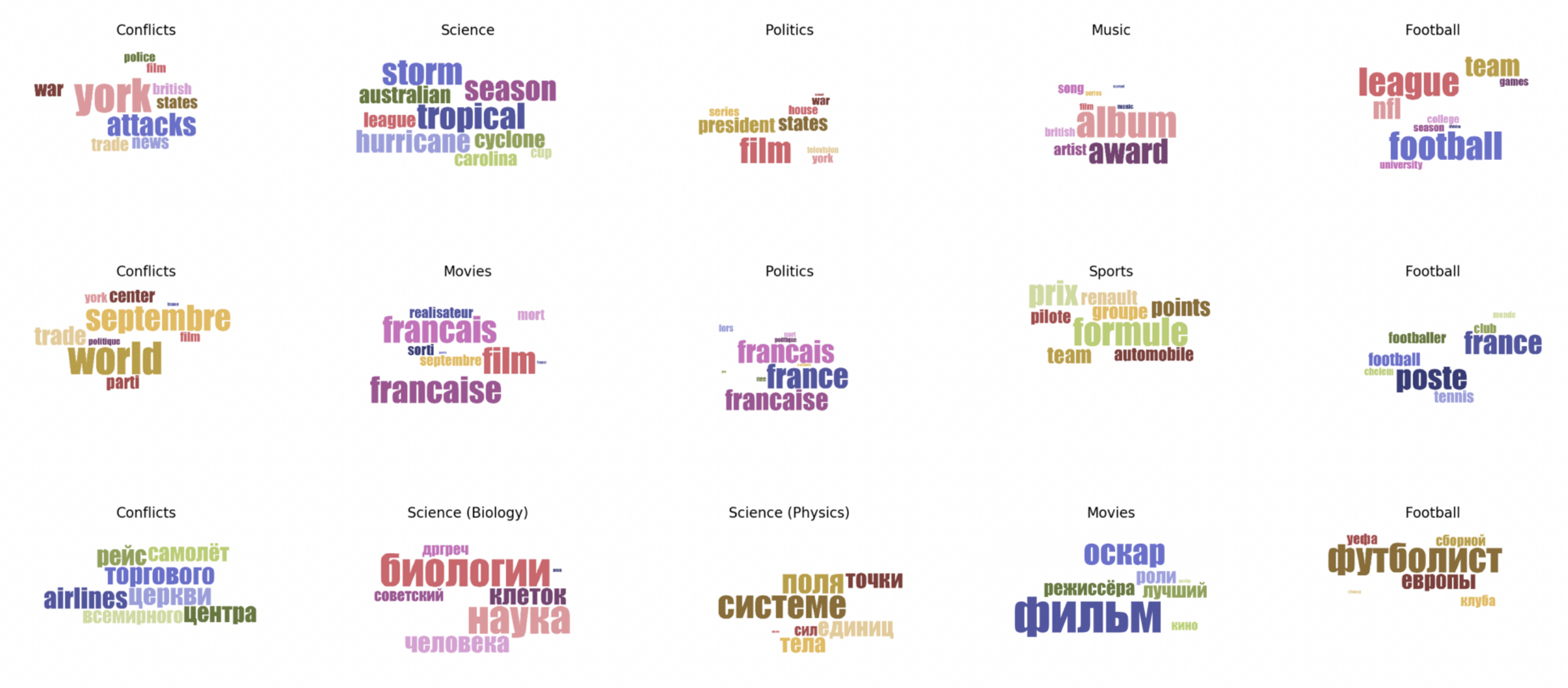}
  \caption{Word-clouds representing the keywords describing the most popular topics of interest in three language editions of Wikipedia (1-16 September 2018). Note that some topics are different across languages since not all of the topics triggered visitors' interest during that period. Top: English. Middle: French. Bottom: Russian.} 
  \label{fig:teaser}
\end{figure*}

\textbf{Step (2): key words extraction.}  In our case, topic modeling is non-trivial since we may have a set of semantically incoherent Wikipedia pages in one cluster. For example, a cluster related to a terrorist attack can comprise pages related to geographic landmarks, politicians, ideology, religion or other attacks that people read to complement information (see~\cite{miz2019anomaly}).

Wikipedia categories found in the trending articles are not appropriate to define a high-level topic for every cluster (too narrow to describe the general idea or topic). Instead, we collected summaries of every article using the Wikipedia API. Then, we used topic modeling and semi-supervised learning to define and assign high-level topics to every article in the extracted sub-network of Wikipedia.
In addition to the text, we take advantage of the graph attributes to provide more context to the detected topics.

We tested 2 methods, Latent Dirichlet Allocation (LDA)~\cite{blei2003latent} and Term Frequency -- Inverse Document Frequency (TF-IDF)~\cite{sparck1972statistical}. Both methods were compared qualitatively and were found to yield similar results. 

For both methods, to improve the performance of the algorithms, we use the features extracted from the graph structure. Vanilla LDA and TF-IDF models demonstrated poorer performance compared to our graph-based alternatives. Since we deal with semantically incoherent documents (details in Step 1), the vanilla LDA model showed poor performance. The detected topics were qualitatively worse than the ones detected using the proposed graph-based alternative.

To train LDA, we pre-process a group of articles and we create the Document-Term Matrix that will be fed into LDA, one document being a concatenation of all articles in a given community. We perform the training in three steps. First, using the entire text, second, with nouns only and third, with the nouns and the adjectives. To look at terms that belong to one part of speech (nouns or adjectives) we used part-of-speech tags from the Penn Treebank Project~\cite{Santorini1990}.

Concerning TF-IDF, we extract $k$ most descriptive words from each group of articles using their TF-IDF scores. However, we modify the TF-IDF method as follows. When computing TF-IDF coefficients, we take advantage of the graph structure of each cluster using the degrees of the nodes to get more context about the cluster. Since the degree of a node is the number of edges connected to it, high degree nodes are more important in semantic sense and give more context related to the topic. We compute the degree $n$ of every Wikipedia article within the sub-network and multiply the counts of every word in this document by $n$. That way, we give more importance to the words in the articles that have high-degree (more central in the cluster) by increasing the TF value while keeping the same IDF value.

\textbf{Step (3): labeling.} 
In order to compare different languages, we select a small set of keywords and their translations for the labeling. In the beginning, we defined 27 specific labels. However, the distribution of the number of samples per class was skewed, which made the classification problem more complicated. To solve this problem, we defined more general categories and limited the number of high-level labels to eight. The labels are \textit{football, sports (other than football), politics, movies, music, conflicts, religion, science, and videogames.} Let us notice that the word "conflicts" is more general than military conflicts. We use it to label a broad range of traumatic events that result in numerous fatalities. Therefore, we also included natural disasters, mass shootings, terrorist attacks, and airplane crashes into this category.  

We then proceed in a semi-supervised manner. Some of the pages are easier to label and we design a few rules to label them automatically. We take advantage of the fact that some articles are homonyms and, to prevent ambiguity, their title contains some useful information in parenthesis such as "album", "actor", or "footballer". For example, all articles that have a title with this pattern "xxx (album)" will be classified as "music". For the remaining unlabelled articles, we use the keywords that we have extracted during the previous step. We define multiple sets of words, where each set corresponds to a label. As an example, the keywords "political", "party", and "republican", represent the topic "politics". The labeling loops through all the summaries of Wikipedia articles, and checks if all these keywords are present in an article’s extract. If the condition holds, it labels the article as "politics". It does the same for all other sets of keywords that represent different labels. 

Let us remark that this could be performed even more efficiently using Wikidata (see Sec.~\ref{sec:discussion} for more information). 

After labeling a subset of Wikipedia articles, we can train a classifier to label the rest. We choose to use a neural network for the labeling task, although in this case, other classifiers perform also well (we also tried SVM, which gave good results).  





In order to classify unlabeled articles, we used a pre-trained BERT model~\cite{devlin2018bert} uncased, with 12 hidden layers and a hidden size of 768 (i.e. the last hidden layer generates a 768-dimensional vector for every word). BERT also stacks multiple attention layers. Attention heads enable the model to capture the relationships between words, giving more weights to some words than to others. This model is trained on two main tasks. First, \emph{Masked Language Modeling} (MLM), where 15\% of the word tokens are masked, and BERT is trained to predict the correct word. Second, \emph{Next Sentence Prediction} (NSP), where the model is fed two sequences A and B and is asked to predict whether B is the next sentence after A. We use the pre-trained model as a feature extractor for the Wikipedia articles. To complete the model, we added a classifier on top of it, consisting of five fully-connected layers that take the features extracted by BERT as input.

\section{Experiments and results}\label{sec:results}

To illustrate each aspect of the method, we demonstrate it from three different points of view. We first take a global view of the trends by looking at the bursts of visits over a 4 month period (16 August to 31 December 2018). We then zoom in to describe concrete examples of the popular topics over a short period of time, 1-16 September 2018. We discuss the associated keywords and related labels. We finish by presenting statistics highlighting similarities and differences between three language editions of Wikipedia.

\textbf{Trends and events during Sep.- Dec. 2018}
On Fig.~\ref{fig:teaser_timeseries}, we depict the most popular trends for 3 languages over the last 4 months of 2018. Notice that more trends were extracted from step (1) but we limited them to the most popular ones in order to get a clear picture. Each colored curve represents the evolution of the popularity of a trend over time. We annotate each peak of the popularity with the title of the most central page of the trend. More precisely, for each cluster (obtained after step (1) of the processing), we select a page having the largest page rank. Vertical red dashed lines highlight remarkable dates where some trends appear synchronously across languages, for example, the 09/11 commemoration or the death of Stan Lee in November. Some trends can be seen in 2 of the 3 languages, such as Formula 1 in French and Russian or the death of George H. W. Bush, in November, in Russia and the US. Some are only seen in one of the languages and are related to the cultural or local events such as the death of a popular Russian singer in September, French activist and politician Nicolas Hulot publishing a book, the death of the former presenter of a popular TV show in France "Nulle Part Ailleurs" or the Miss France contest. We can also remark that many of the trends are related to entertainment such as movies, music or sports. All of the trending topics have been covered by the media, showing their massive influence on Wikipedia readers. 

\begin{figure}
  \includegraphics[width=8.5cm, trim={5cm 0cm 5.5cm 0cm}, clip]{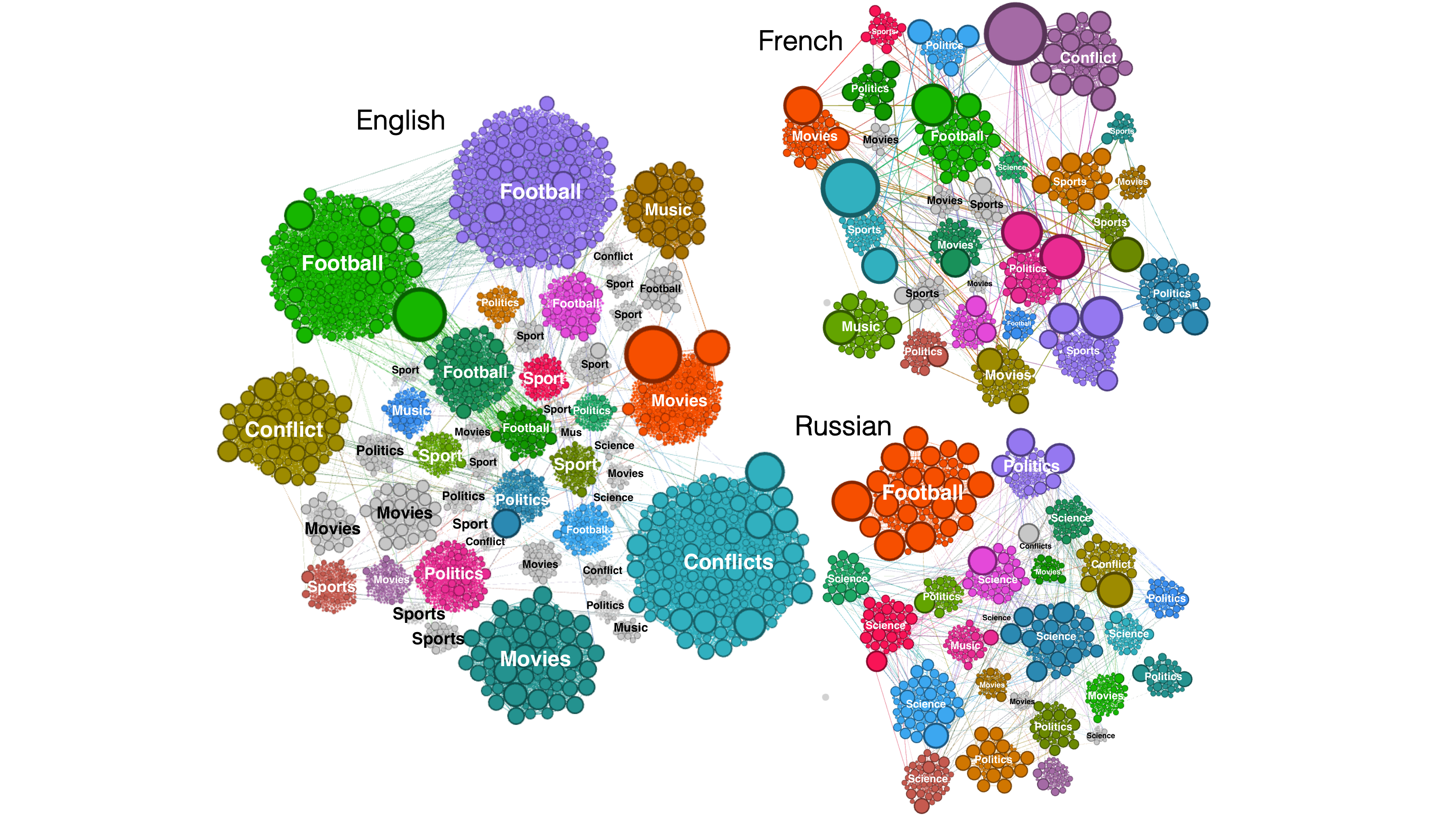}
  \caption{Subnetworks of trending Wikipedia topics in three language editions (1-16 September 2018). Nodes are Wikipedia pages having a peak of visits somewhere within these 2 weeks. Edges are hyperlinks connecting these pages. Connections are strengthened between pages if the number of visits peaks at the same time. As a consequence, each cluster is a trend related to some event that drew the interest of Wikipedia readers at that moment. The clusters have been automatically labeled. Interactive visualizations showing clusters of trending pages in each language edition are available online \href{https://wiki-insights.epfl.ch/wikitrends}{https://wiki-insights.epfl.ch/wikitrends}
  }
  \label{fig:graphs}
\end{figure}

Most of the bursts have a similar duration of a few days, with a similar peak shape across all languages. This indicates a common behavior, independent of the culture or language, about the trends and the time span of readers' interest in a topic.

\textbf{1-16 September 2018.} We focus on five most popular trends in each language in order to give an example of the results of the keyword extraction. For each language, we get the distribution of keywords presented in Fig.~\ref{fig:teaser}. Each word cloud represents keywords describing a topic. 

First, let us look at shared topics. We can see a common topic in three Wikipedias. The attacks of September 11 in 2001 on The Twin Towers of the World Trade Center. It is represented by words such as "attacks", "September", "center", "trade", "world".

We can also see unique topics in each language. For example, there is a weather-related topic in English Wikipedia, which could be explained by the series of hurricanes that caused considerable damage in North and South Carolinas, such as Hurricane Florence in September 2018. This topic appears neither in Russian nor in French Wikipedia. This also holds for Hurricane Michael visible only in the English part (October) in Fig.~\ref{fig:teaser_timeseries}. Another example is two science-related topics in Russian Wikipedia focusing on Biology and Physics. Finally, we can see a topic related to Formula 1 only in French Wikipedia.

\begin{figure}
  \includegraphics[width=8.5cm, trim={2cm 0cm 2cm 0cm}, clip]{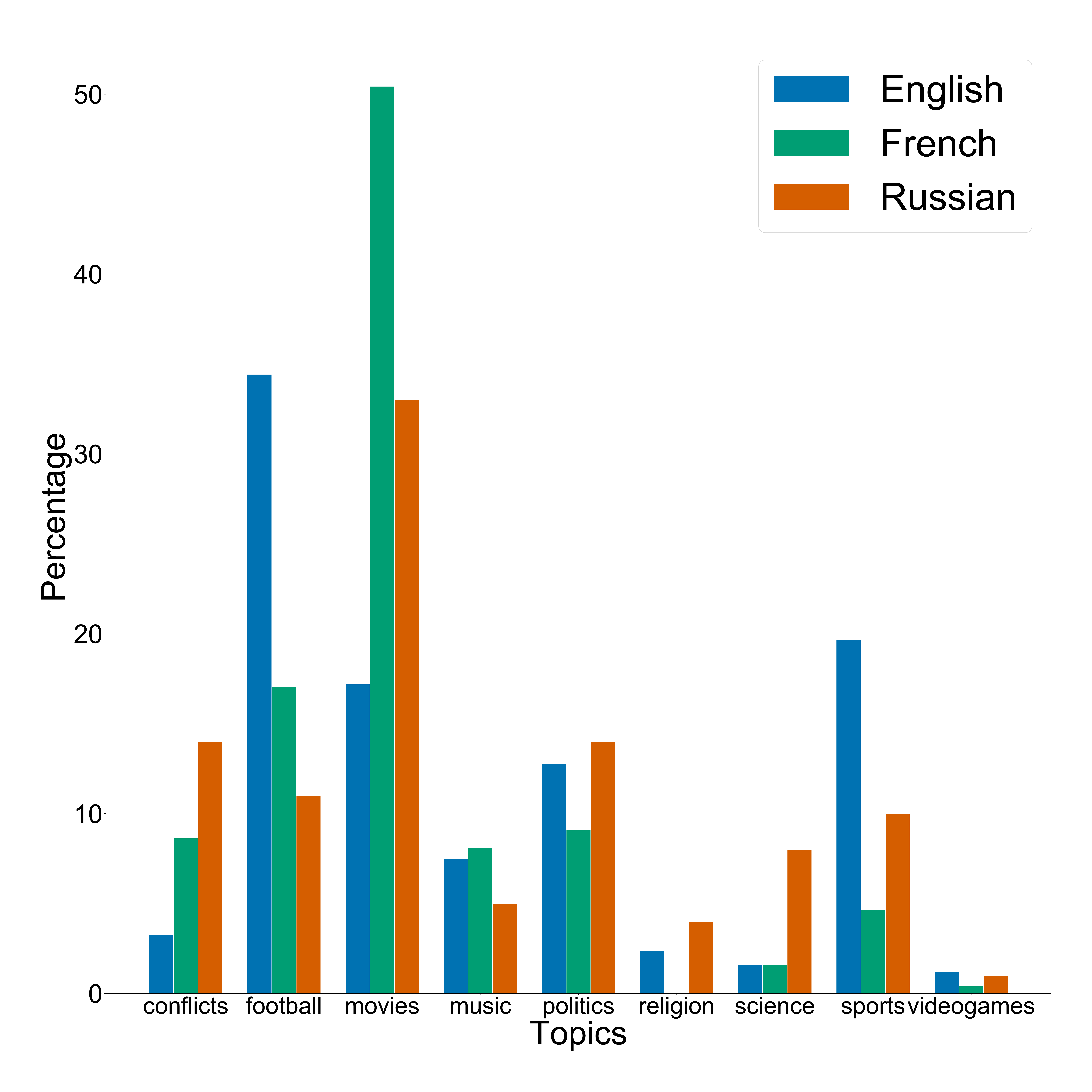}
  \caption{Distribution of the most popular topics across languages (16 August -- 31 December 2018)}
  \label{fig:distributions}
\end{figure}

The network of clusters for each language extracted during this period of time is shown in Fig. \ref{fig:graphs}. 
These networks are the output of step (1). The figure illustrates the Clusters of highly connected pages that are more or less connected to each other. Each cluster is associated with a different trend that has been automatically labeled. Two clusters may have the same label but two different topics. For example, one of the large Football clusters in the English networks corresponds to European football (UEFA) and the other to American football (NFL). This is a reminder that cultural differences also exist within a language, especially if this language is shared across multiple continents. These graph representations shed some light on the curiosity of people who can read many different pages related to a single topic. For Hurricane Florence for example, readers do not only visit the page of the hurricane written and edited live, they also follow hyperlinks to pages of the affected cities, to pages of previous hurricanes or similar natural disasters, to pages defining a cyclone or providing a scientific explanation of this phenomenon. More than a hundred different pages have encountered a burst of visits during this event. 

The size of the clusters can also reveal cultural differences. The spike of visitors on the page dedicated to Stan Lee was very high in the three languages, shortly after his death. However, the cluster size was much smaller in the French and Russian versions. English speakers massively explored his work through the hyperlinks to pages about the comics, the movies based on them and the starring movie actors. Most of the francophone readers focused only on his biography, despite the existence of many hyperlinks pointing towards the main heroes of his comics or the movies adapted from them on his French Wikipedia page. On the French Wikipedia cluster, apart from Stan Lee, only the pages of Magneto, Jack Kirby, Steve Ditko, Larry Lieber, POW! entertainment, CNN and the expression "Excelsior" are present in the cluster.

\textbf{Global statistics over 4 months.} Figure \ref{fig:distributions} shows the distribution of the most popular topics across the different editions of Wikipedia over the period of the last four months of 2018. We can see that while English-speaking readers prefer topics related to football and general sports, the most popular content among French and Russian-speaking visitors is mostly related to the movie and TV industry. The other topics related to entertainment, music and video games, get almost equal attention in all languages. Politics is equally popular in the 3 edition of Wikipedia, while the content about geopolitical issues (natural disasters, foreign affairs or conflicts) appears to draw more interest from French and even more from Russian. Science is among the most popular topics for Russian-speaking readers while being equally low in French and English. French-speaking readers have the lowest interest in religion, in contrast to English and Russian ones. English-speakers appear to be more interested in religion than in science. We compare our results to the ones of~\cite{lemmerich2019world} in the next section.

\section{Discussion}\label{sec:discussion}

\textbf{Why are some topics different across languages?} 
The difference could be attributed to multiple reasons and we only try to cover a few of them. First, \textit{media coverage} and Wikipedia's featured articles that appear on the main page of Wikipedia. The trends are a mass phenomena that are related to important topics and, as such, are well covered by the media, as pointed out in Sec.~\ref{sec:results}. The question is whether the media increases or inflates the interest of people in the trends. The authors of~\cite{singer2017we,lemmerich2019world} give a positive answer: on average around 25\% of readers are motivated by media coverage, it even reaches 30\% for the English and Russian version. Some of the trends we captured suggest a positive answer as well. For example, when a famous person dies, particularly in show business, media often broadcasts some of their works, be it movies (death of Stan Lee), TV shows (death of P. Gildas, "Nulle part Ailleurs" event on the French Wikipedia in October) or music (death of J. Kobzon event on the Russian Wikipedia in September). This would, in turn, increase the curiosity of people about the person and their works. Besides events getting more attention from their media coverage, we may assume some trends to appear exclusively due to media providers that produce and advertise their TV shows, such as Miss Universe, Miss France, or even Emmy awards visible in Fig.~\ref{fig:teaser_timeseries}. 

Some trends may be less influenced by the media. For example, sports events do not mainly rely on the media to attract the interest of fans. In this case, the reader's motivation could fall into the category of "conversation" or "event" as described in~\cite{singer2017we,lemmerich2019world}. In that study, these types of motivation have also high scores on average: 24\% for "conversation" and 17\% for "event" (motivation triggered by the event itself) and similar values for the English and Russian editions. In our results, sports (including football) is the second most popular topic after movies and even the first one in the English Wikipedia.

It is important to remark that if the media alone were the only driving force of the readers' interests on some Wikipedia topics, the trends would have had a different shape. Their clusters would have been made of a single or a few pages, as people would have gone to the page covering the event and left Wikipedia after the first read. Indeed, this phenomenon is common and can be observed on Wikipedia pages that are highlighted by some popular websites such as Wikipedia's or Google's front pages. Although, in our study, we filter out pages that have a single spike and only select clusters of connected pages with a correlated increase of visits. Again, referring to~\cite{singer2017we,lemmerich2019world}, the main source of motivation of Wikipedia readers is "intrinsic learning" (except for the English version where it comes second) with 37\% on average. Readers, motivated by the media, may visit a page but they stay and follow hyperlinks on Wikipedia because they are motivated by \emph{intrinsic learning}. This is demonstrated by the very existence of the clusters of pages that we capture in our study. Moreover, many pages in each cluster bring complementary information that is not covered by the media or is not directly related to the event itself. For example, in the Hurricane Florence cluster, we can see some pages that list past hurricanes and pages related to meteorology showing the importance of hyperlinks for intrinsic learning.

Here, Wikipedia's structure plays a significant role since the readers rely on inter-article links that lead to the related content. Indeed, readers follow hyperlinks and are in search of more information than just basic facts about the trends. This behavior is equally shared across languages although clusters are smaller in the French and Russian versions than in the English one. We assume this is due to the smaller size of the Wikipedia network and the smaller number of people reading Wikipedia in other languages. The English version of Wikipedia contains around 6 million pages while the French one contains around 2 million and the Russian around 1.5 million. 

Second, \textit{geographic proximity}. This is especially apparent when we consider natural disasters. For example, the hurricane in North Carolina did not spark interest among the French- and Russian-speaking readers and appeared only in the English edition of Wikipedia. Although, we can see that some outrageous traumatic events trigger memories among the readers in all languages. We can see the topic related to the 9/11 attacks as a supporting example. 


\begin{figure}
  \includegraphics[width=8.5cm, trim={1cm 0cm 2cm 0.9cm}, clip]{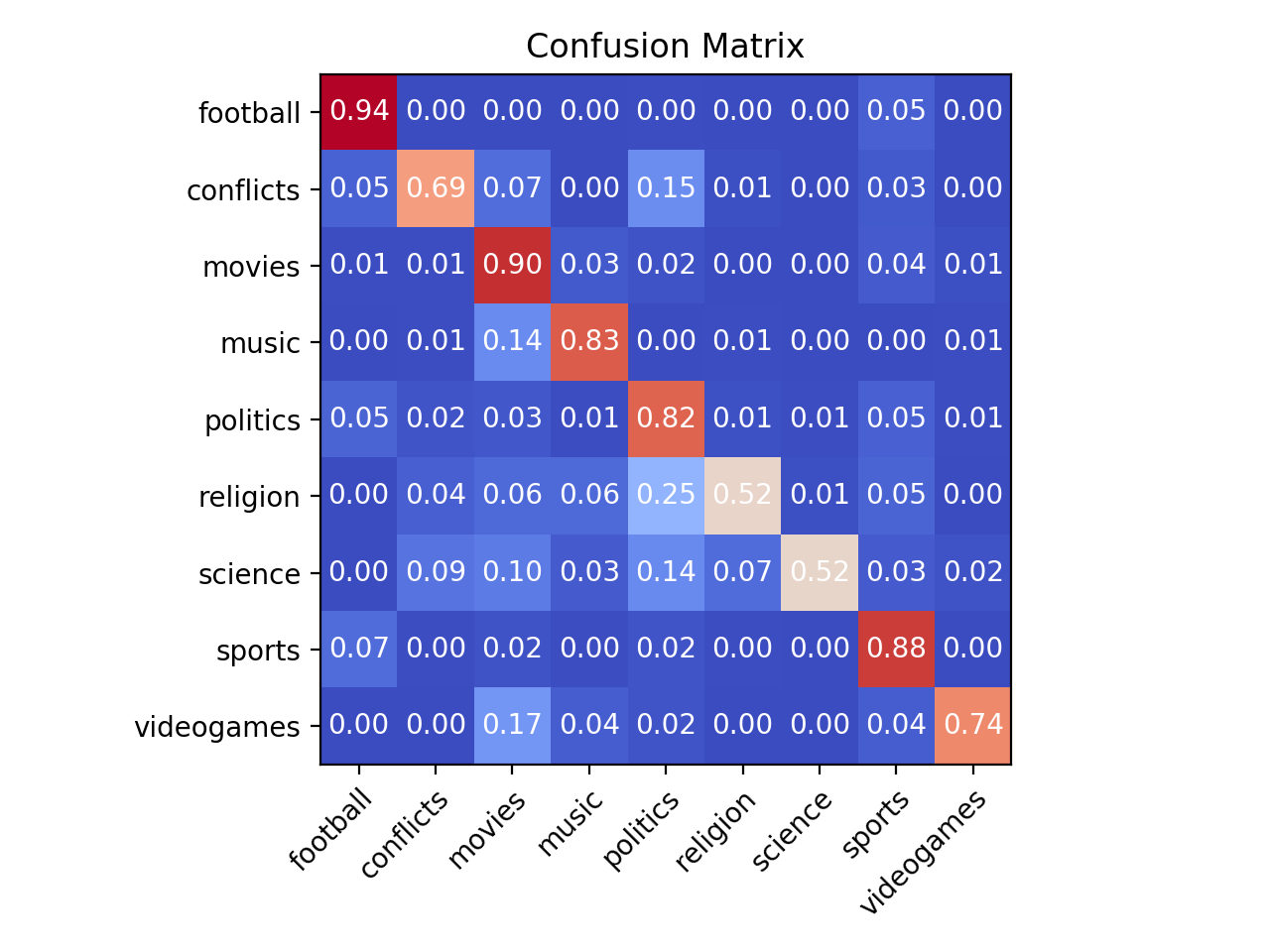}
  \caption{Confusion matrix of BERT classification model.}
  \label{fig:confusion_matrix}
\end{figure}

Finally, \textit{cultural differences}. This difference is especially vivid when we look at sports. Readers of English Wikipedia (mostly dominated by the readers from the USA) tend to be interested in NFL championships, while the French or Russian-speaking readers do not express as much excitement about this topic and prefer European football championships (soccer). However, some sports such as golf or tennis are equally interesting to all groups of readers. This cultural influence can also be seen in music- and movies-related trends that are different across languages. Moreover,~\cite{lemmerich2019world} also shows cultural differences in the motivation of the reader. For example, the "intrinsic learning" motivation between western and eastern cultures is different, with higher values for eastern people such as Russian-speaking readers. This is confirmed in our study where topics related to science have a higher number of clusters in the Russian edition of Wikipedia (Fig.~\ref{fig:distributions}). An example can be seen in Fig.~\ref{fig:teaser_timeseries} with the "Soyuz spacecraft" peak in the Russian version in October.  It is the only science-related trend among the top trends in this timeline. Also, we can see this in Fig.~\ref{fig:graphs} where clusters related to science prevail in the Russian edition's graph.

\textbf{Limitations of automated labeling.} Automated labeling is not 100\% accurate. We analyze the errors made by the classification model and show the results in Fig. \ref{fig:confusion_matrix} and in Table \ref{tab:classificaion_metrics}. Qualitative analysis of the errors shows that they are not significant in terms of semantics or meaning of the misclassified topics. Errors are mostly caused by the overlap of vocabularies across different topics.
First, let us look at the category "Football". In the BERT-based classification model, the model confused "Football" with "Sports". Similar behavior is noticed for the label "Sports". 
Indeed, "Football" as "Sports" classes have similar meaning so it is natural even for humans to misclassify the two. Similarly, music and movies are mixed as well as politics and conflicts or religion as they may involve the same people. This indicates that the classification would improve if we choose more detailed topics that are more specific and had a more balanced dataset rather than training with more data.

\textbf{Alternative sources of data for labeling.} We have also tested Wikidata as an alternative data source for automated labeling instead of Wikipedia article summaries. Wikidata is a structured knowledge base focused on items that represent high-level topics, concepts, and categories. It demonstrated comparable performance and produced relevant keywords allowing to infer high-level topics for the clusters by collecting properties for each page and extracting the topic from the resulting high-level summary with LDA.  Due to the structure of the knowledge base, there is a major positive aspect of using Wikidata for labeling and topic detection purposes. Wikidata properties and concepts are identified by a unique code, with direct relationships to their equivalent in all the Wikipedia languages. Therefore, there is no need to deal with different languages when running the topic detection pipeline. Also, this reduces the need for experts in every studied language to assess the quality of the topic detection, hence scaling the study to more languages would become much easier. The only drawback is that querying Wikidata API is relatively slow and when it comes to collecting data for thousands of pages, this step quickly becomes a bottleneck.\footnote{To test this data source, we deployed the entire Wikidata inside a local MongoDB instance. As of January 2020, the size of the deployed Wikidata knowledge base is around 220 GB.} Wikidata word extraction gave promising results with relevant keywords for the clusters, and we plan to replace article summaries by Wikidata in our future work. For all these reasons, we recommend using Wikidata for labeling and topic detection in future studies based on the present work.

\begin{table}[t!]
\begin{center}
\caption{Classification metrics \label{tab:classificaion_metrics}}
\begin{tabular}{crrrr}
\toprule
 & {\bf Precision} & {\bf Recall} & {\bf F1} & {\bf Support}\\ 
\midrule
Football & 0.93 & 0.95 & 0.94 & 1359 \\ 
Conflicts & 0.75 & 0.69 & 0.72 & 112\\ 
Movies & 0.86 & 0.89 & 0.88 & 646\\ 
Music & 0.84 & 0.85 & 0.84 & 288\\ 
Politics & 0.82 & 0.81 & 0.82 & 520\\ 
Religion & 0.75 & 0.56 & 0.64 & 90\\ 
Science &0.79 & 0.60 & 0.68 & 70\\ 
Sports & 0.84 & 0.85 & 0.85 & 751\\ 
Videogames & 0.79 & 0.70 & 0.74 & 47\\ 
\bottomrule
Accuracy &   &   & 0.87 & 3883\\ 
Macro AVG & 0.82 & 0.77 & 0.79 & 3883\\ 
Weighted AVG & 0.87 & 0.87 & 0.87 & 3883\\ 
\bottomrule
\end{tabular}
\end{center}
\end{table}


\section{Conclusions}
In this study, we have compared the most trending topics based on the viewership statistics of English, French and Russian Wikipedia editions. The method is able to follow and capture the dynamic evolution of topics and interests of Wikipedia readers, for any language. We have analyzed the distribution of visitor interests over the period of September-December 2018 for the three languages and highlighted the main differences and similarities in the interests among the readers of Wikipedia.

In future work, firstly we would like to extend our study to other languages, taking advantage of the available pre-trained multilingual models of BERT. Alternatively, a language-independent automatic labeling, possibly based on Wikipedia categories, could be designed for the application to rare languages. BERT might be limited in this configuration. Secondly, it would be interesting to extend the time period to several years and introduce additional keywords for the classification.

\begin{acks}
We would like to thank the anonymous reviewers for their insightful comments and constructive suggestions that helped to improve the quality of this paper.
\end{acks}

\balance

\bibliographystyle{ACM-Reference-Format}
\bibliography{main}

\end{document}